\documentclass[twocolumn]{aastex61}
\usepackage{color}
\usepackage{hyperref}
\usepackage{latexsym}
\usepackage{graphicx,epsf,epsfig}% Include figure files
\usepackage{amssymb,amsmath}   % for math
\usepackage{cancel}
\usepackage{soul}

\defcitealias{2017ApJ...846..125B}{Paper~I}
\defcitealias{BeS_2018}{Paper~II}

\submitjournal{ApJ}
\begin{document}

\title{The discreteness-driven relaxation of collisionless gravitating
  systems: entropy evolution and the Nyquist-Shannon theorem}

\author{Leandro {Beraldo
    e Silva}} \affiliation{Universidade de S\~ao Paulo, Instituto de
  Astronomia, Geof\'isica e Ci\^encias Atmosf\'ericas, Departamento de
  Astronomia, CEP 05508-090, S\~ao Paulo, SP, Brasil}
\affiliation{Department of Astronomy, University of Michigan, Ann
  Arbor, MI 48109, USA} \email{lbs@usp.br} \author{Walter {de Siqueira
    Pedra}} \affiliation{Universidade de S\~ao Paulo, Instituto de
  F\'isica, Departamento de F\'{\i}sica Matem\'atica, CP 66318, CEP
  05314-970, S\~ao Paulo, SP, Brasil} \author{Monica Valluri}
\affiliation{Department of Astronomy, University of Michigan, Ann
  Arbor, MI 48109, USA} \date{\today}

%\maketitle
\begin{abstract}
  The time irreversibility and fast relaxation of collapsing $N$-body
  gravitating systems (as opposed to the time reversibility of the
  equations of motion for individual stars or particles) are
  traditionally attributed to information loss due to coarse-graining
  in the observation. We show that this subjective element is not
  necessary once one takes into consideration the fundamental fact
  that these systems are discrete, i.e. composed of a finite number
  $N$ of stars or particles. We show that a connection can be made
  between entropy estimates for discrete systems and the
  Nyquist-Shannon sampling criterion. Specifically, given a sample
  with $N$ points in a space of $d$ dimensions, the Nyquist-Shannon
  criterion constrains the size of the smallest structures defined by
  a function in the continuum that can be uniquely associated with the
  discrete sample. When applied to an $N$-body system, this theorem
  sets a lower limit to the size of phase-space structures (in the
  continuum) that can be resolved in the discrete data. As a
  consequence, the finite $N$ system tends to a uniform distribution
  after a relaxation time that typically scales as $N^{1/d}$. This
  provides an explanation for the fast achievement of a stationary
  state in collapsing $N$-body gravitating systems such as galaxies
  and star clusters, without the need to advocate for the subjective
  effect of coarse-graining.
\end{abstract}

%\pacs{}
\section{Introduction}
An important question in the study of collisionless $N$-body
gravitating systems is how to reconcile the time irreversibility of
the fast relaxation of a collapsing structure with the reversible
character of the equations of motion for the individual stars or
particles. In other words, how to reconcile this irreversible
relaxation with the time reversibility of the Vlasov equation (assumed
to describe the kinetic evolution). The most common solution to this
apparent paradox \citep[``the fundamental paradox of stellar
dynamics'', according to][]{1965dss..book.....O} is to attribute this
time irreversibility to information loss in a ``coarse-grained''
observation \citep[][]{LyndenBell_1967,
  2014PhR...535....1L}. According to this view, during the dynamical
evolution of the system, the distribution of particles in phase-space
progressively develops finer and finer structures (i.e., filaments)
that after some time can no longer be detected by the observing
device, which only measures averaged (``coarse-grained'') quantities.

A fundamental problem with this solution, however, is that it
introduces a subjective element, making the relaxation phenomenon
dependent on the observational precision
\citep{1965AmJPh..33..391J}. In this paper, we provide an alternative
scenario without this subjective element. We use rigorous recipes to
estimate the entropy of a discrete sample \citep[see][]{Joe1989,
  Beirlant1997a, Leonenko_2008, biau2015lectures} and show that its
evolution is connected to the celebrated Nyquist-Shannon sampling
theorem of signal theory and image processing.
%\citep[][]{5055024, 1697831}.

For systems with $N$ particles evolving in a phase-space of dimension
$d$, we derive a relaxation time that scales typically as
$N^{1/d}$. Once one recognizes as a fundamental fact that
gravitational systems such as galaxies and star clusters are
finite-$N$ systems, as opposed to the theoretical limit to the
continuum (${N\rightarrow \infty}$), this time scale is naturally seen
as a real relaxation time. Our analysis provides a theoretical
explanation for the power law $N$-dependencies of the relaxation time
obtained by \cite{2017JSMTE..04.4001P} for long-range interacting
systems in $d=2$ and by \cite{BeS_2018} for ensembles of orbits
integrated in triaxial gravitational potentials ($d=6$). Finally,
together with the analysis of \cite{2017ApJ...846..125B} (hereafter
{\bf Paper I}) and \cite{BeS_2018} (hereafter {\bf Paper II}), this
discreteness effect is shown to be the main mechanism for the fast (in
a few dynamical time scales) collisionless relaxation of
non-equilibrium $N$-body gravitating systems.

In \S~\ref{sec:baby_model} we show analytically that the finest
phase-space structures of an ensemble of free particles are expected
to develop linearly in time. In \S~\ref{sec:n-s} we briefly introduce
the Nyquist-Shannon theorem, relating the inverse size of these finest
phase-space structures (bandwidth) with the size $N$ of a discrete
sample. In \S~\ref{sec:toy_model} we develop a non-dynamical toy model
with known values of the bandwidth and in \S~\ref{sec:entropy} we
introduce the entropy estimator, showing that the estimates applied to
the toy model agree with the Nyquist-Shannon criterion. In
\S~\ref{sec:relax_dynamical_systems} we apply the entropy estimator to
the study of the relaxation of orbit ensembles integrated in a Plummer
potential, deriving the typical relaxation time. Finally,
\S~\ref{sec:summary} summarizes our results.

\section{The simplest dynamical model}
\label{sec:baby_model}
In this section we consider the simple example of an ensemble of free
particles to show how the phase-space structures are expected to
evolve in time. In particular, in order to make contact with the
Nyquist-Shannon theorem in the next sections, we are interested in the
time evolution of the finest phase-space structures, i.e. of the
largest structures (the bandwidth $K$) in Fourier space of the
distribution function.

Consider an ensemble of particles initially distributed according to a
Gaussian in a $d$-dimensional phase-space:
\begin{equation}
  \label{eq:f_0}
  f_0(\vec{x},\vec{v}) =
  \frac{1}{\left(2\pi\right)^{d/2}}\exp{\left(-\frac{\vec{x}^2 + \vec{v}^2}{2}\right)},
\end{equation}
where $\vec{x}$ and $\vec{v}$ are conveniently normalized
dimensionless position and velocity. If no forces act on the
particles, at a time $t$ this distribution evolves to
\begin{equation}
  \label{eq:f_t}
  f_t(\vec{x},\vec{v}) =
  \frac{1}{\left(2\pi\right)^{d/2}}\exp{\left[-\frac{\left(\vec{x} - \vec{v}t\right)^2 + \vec{v}^2}{2}\right]}.
\end{equation}
Note that if Eq.~(\ref{eq:f_t}) is seen as characterizing the
macroscopic state, i.e. considering the system as a whole, as is
normally the case when referring to a (probability) distribution
function, it associates a number to each point $(\vec{x},\vec{v})$ in
a continuous $d$-dimensional domain for every time $t$. In this way,
when describing the evolution of a finite-$N$ system, it necessarily
extrapolates the information available in a discrete sample,
implicitly assuming that this extrapolation is always physically
meaningful.

Taking the Fourier transform of Eq.~(\ref{eq:f_t}), one gets (see the
appendix for more general and detailed calculation)
\begin{align*}
  \label{eq:fourier_1}
  \hat{f}_{t}(\hat{x},\hat{v}) &\equiv \frac{1}{(2\pi )^{d/2}}\int
  \mathrm{d}\vec{x}\int \mathrm{d}\vec{v} \,
  e^{\left[-\frac{(\vec{x}-\vec{v}t)^{2}+\vec{v}^{2}}{2}\right]} e^{-i
    (\hat{x}\cdot \vec{x} + \hat{v}\cdot\vec{v})} \\
  &= \exp\left[-\frac{(t^{2}+1)\hat{x}^{2}+2t \hat{x}\cdot\hat{v} +
    \hat{v}^{2}}{2}\right],
%   &= \exp \left( -\frac{1}{2}\hat{X}^T {\bf A} \hat{X}\right),
\end{align*}
where ($\hat{x},\hat{v}$) are the respective wavevectors, or
frequencies, in Fourier space. We are interested in determining the
largest characteristic scales, i.e. the bandwidth $K$, in Fourier
space and a rough estimate can be made as
\begin{eqnarray*}
  K^2(t) &\approx& \langle \hat{x}^2 + \hat{v}^2\rangle = \frac{\int\mathrm{d}\hat{x}\int\mathrm{d}
     \hat{v}(\hat{x}^{2}+\hat{v}^{2})\hat{f}_{t}(\hat{x},\hat{v})}{\int\mathrm{d}\hat{x}\int\mathrm{d}\hat{v}\hat{f}_{t}(\hat{x},\hat{v})}\\
%         &=& \int\mathrm{d}\hat{X}^{d}\hat{X}^2
%            \exp \left( -\frac{1}{2}\hat{X}^T {\bf A} \hat{X}\right)\\
%         &=& \sqrt{\frac{(2\pi)^d}{(\det {\bf A})^{d/2}}}\cdot
%             \sum_{i=1}^d \left( {\bf A}^{-1}\right)_{ii}\\
         &=&(2+t^{2})\text{ }.
\end{eqnarray*}
This already shows that at large times the bandwidth $K$ is expected
to grow linearly with time for this simple model. A different estimate
takes into accout that the characteristic scales of
$\hat{f}_t(\hat{x},\hat{v})$ can be associated with the dispersions in
two orthogonal directions obtained by some rotation of the
$(\hat{x},\hat{v})$ axes. Since we are interested in the largest
characteristic scales, we make a rotation such that one of the new
axis points in the direction of the largest dispersion. This is
similar to the so-called principal component analysis and is made in
the appendix, where we similarly conclude that $K(t)\propto t$ for
large times.

\section{Nyquist-Shannon theorem}
\label{sec:n-s}
According to the Nyquist-Shannon (hereafter {\bf N-S}) theorem, a
function in the continuum can be recovered from its discrete sampling
whenever the sampling rate is at least twice the bandwidth $K$ in
Fourier space of the function (N-S criterion). In the original proofs
of \cite{5055024} and \cite{1697831}, the sampling was assumed to be
uniform, but their result was extended, later on, to the case of
nonuniform samplings, in which the sampling rate is to be understood
as the average sampling rate \citep{landau1967}. This sufficient
condition to recover a function in the continuum from some discrete
sampling of it is also known to be necessary, in general. According to
this theorem, in order to exactly reconstruct a function in a
$d$-dimensional continuum (i.e. in a continuous domain) from a
discrete sample, the number of sampling points must be
$N \gtrsim K^d$, where $K$ is the largest characteristic frequency of
the function i.e. its bandwidth in frequency/Fourier
space. Conversely, given an arbitrary discrete sample with $N$ points,
the theorem states that only functions with bandwidth
\begin{equation}
  \label{eq:nyquist}
  K \lesssim N^{1/d},
\end{equation}
i.e. with structures not finer than given by Eq.~\eqref{eq:nyquist},
can be uniquely associated to the sample. Functions with a larger
bandwidth, i.e. with finer structures, contain extra information not
equivalent to that in the sample.

\section{Toy model}
\label{sec:toy_model}
In this section we introduce a (non-dynamical) toy model based on a
simple probability distribution function for $d$ independent variables
\begin{equation}
  \label{eq:f_d_dim}
  f(x_1,...,x_d)=F(x_1)...F(x_d),
\end{equation}
with
\begin{equation}
  \label{eq:toy_model}
  F(x) = \frac{1}{A}\left\{ 1 + \sum_{m=1}^n \left[a_m\cos(2\pi mkx) + b_m\sin(2\pi mkx)\right]\right\}
\end{equation}
for $-1/2\leq x \leq 1/2$ and $F(x)=0$ otherwise. Here $k,m,n$ are
natural numbers and $A$ is a normalization constant. $A=2$ if $k=0$
and $A=1$ when $k>0$. Note that the case $k=0$ corresponds to a
uniform distribution. In order to ensure that $F(x)$ is non-negative,
the coefficients $a_{m} $, $b_{m}$ are real numbers such that
${\sum_{n}\sqrt{a_{n}^{2}+b_{n}^{2}}\leq 1}$.

In the examples discussed below, for fixed $n>0$ we set $b_{m}=0$ for
all $m$ and $a_{m}=n^{-1}$ for $m\leq n$, with $a_{m}=0$ for
${m > n}$.  We denote by $f^{(n)}$ the distributions corresponding to
these choices of ${a_{m},b_{m}}$. There is nothing special about this
choice, except that the coefficients vanish for ${m>n}$. Other choices
with this property lead to essentially the same results. For this
model the bandwidth is ${K=n\times k}$. Note that this toy model has
no \textit{a priori} dynamical interpretation or time
evolution. However, the parameter $k$ can be seen as analogous to time
$t$, with larger values introducing finer structures in the ``phase
space'' and a linear growth of the bandwidth $K$.

It is important to emphasize that, on the one hand, the imposition of
a function in the continuum restricts the number of sampling points
necessary to correctly recover the function. On the other hand, in
$N$-body gravitating systems what we are given \emph{a priori} is an
arbitrary sample of size $N$. In this case, the N-S theorem imposes
restrictions on the distribution function in the continuum that can be
used to describe the $N$-body system and the equation driving its
kinetic evolution. This point is discussed in
\S~\ref{sec:relax_dynamical_systems}.

\begin{figure*}[htbp]
  \centering
  \includegraphics[scale=0.25]{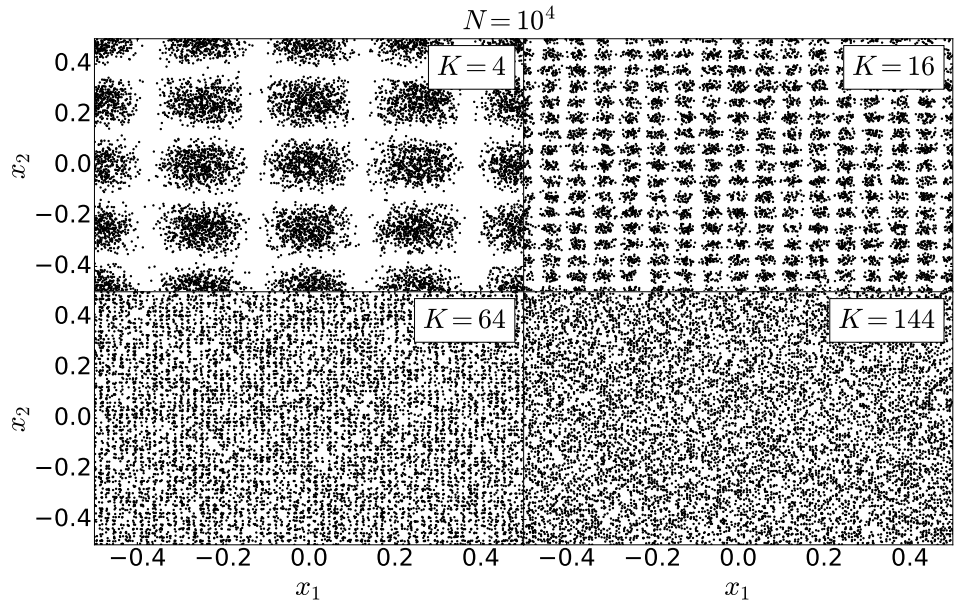}
  \caption{Discrete samples of the toy model
    $f(x_1,x_2)=F(x_1)F(x_2)$, with $F(x)$ defined by
    Eq.~\eqref{eq:toy_model} with $n=1$, using $N=10^4$ points in each
    panel. In agreement with the N-S criterion,
    Eq.~\eqref{eq:nyquist}, for $K\lesssim \sqrt{N}$ the sample is
    able to qualitatively reproduce the features of the generating
    function, while for $K \gtrsim \sqrt{N}$ it resembles a uniform
    distribution.}
  \label{fig:coords_S_2D}
\end{figure*}

Fig. \ref{fig:coords_S_2D} shows samples of this model,
Eqs. (\ref{eq:f_d_dim})-(\ref{eq:toy_model}), with $n=1$ in $d=2$ with
$N=10^4$ points generated with the acceptance-rejection method for
different $k$ values.
%, using $N=10^4$ points.
For small $K$ values, the structure of the distribution function is
well captured by the sample. Larger values of $K$ are associated with
finer structures in the ``phase-space''. Beyond a critical value,
these fine structures cannot be any longer resolved by the sample,
which looks like the one for a uniform distribution. This
qualitatively illustrates the content of the N-S theorem.

\section{Entropy estimates}
\label{sec:entropy}
For a quantitative analysis, we generate samples of the distribution
given by Eqs.~\eqref{eq:f_d_dim}-\eqref{eq:toy_model} for different
values of parameters $k$ and $n$, in different dimensions $d$ and for
different numbers $N$ of points. We then estimate the entropy
associated with each one of these samples as follows: recall first
that the Shannon entropy of the distribution $f(x_1,...,x_d)$ is
defined as
\begin{equation}
  \label{eq:S_def}
  S(k) \equiv -\int f\ln f \,dx_1...\,dx_d.
\end{equation}
Given a sample of points in $d$ dimensions, distributed according to
$f$, this entropy can be estimated as
\begin{equation}
  \label{eq:S_estimate_0}
  \hat{S}(k) = -\frac{1}{N}\sum_{i=1}^N \ln\hat{f}_i,
\end{equation}
where the integral in Eq.~\eqref{eq:S_def} is translated into a sum
over the $N$ sampling points.
%The distribution function at point $i$
%can be estimated $\hat{f}_i$, the estimate of the
%distribution function $f$ at the position of each particle
%$i$.
Eq. \eqref{eq:S_estimate_0} converges to Eq. \eqref{eq:S_def} for
$N\rightarrow \infty$ when we estimate $\hat{f}_i$ with at least two
methods \citep[][]{Joe1989, Beirlant1997a, Leonenko_2008,
  biau2015lectures}: the nearest neighbor and the kernel method. A
study of the evolution of $N$-body self-gravitating systems has shown
that both methods provide very similar entropy estimates -- see
\citetalias{2017ApJ...846..125B}.

In the nearest neighbor method, the distribution function $f$ at the
point $\vec{x}_i=(x_1,...,x_d)$ is estimated as the number of points
inside a hyper-sphere of radius $D_{in}$ (the distance from point $i$
to its nearest neighbor $n$) divided by its volume. With normalization
factors,
\begin{equation}
  \label{eq:f_NN}
  \hat{f}_i = \frac{1}{(N-1)e^\gamma V_d D_{in}^d}
\end{equation}
\citep[see][]{Leonenko_2008}, where $\gamma \approx 0.57722$ is the
Euler-Mascheroni constant, $V_d = \pi^{d/2}/\Gamma(d/2 + 1)$ and
%\begin{equation}
%  \label{eq:dist_phase_space}
$D_{in} = \sqrt{(\vec{x}_i - \vec{x}_n)^2}$.
%\end{equation}
For the identification of the nearest neighbors we use the
\emph{Approximate Nearest Neighbor} (ANN) method
\cite{Arya:1998:OAA:293347.293348}\footnote{Available at
  {www.cs.umd.edu/$\sim$mount/ANN/}. A slightly different version,
  allowing searches in parallel, was developed by Andreas Girgensohn
  and kindly provided by David Mount.}, which is based on a kd-tree
algorithm \cite{Friedman:1977:AFB:355744.355745}. The algorithm allows
one to optimize the search by approximating the nearest neighbor, but
we use it without any approximation, identifying the exact nearest
neighbor.

\begin{figure*}[htbp]
  \centering
  %\epsscale{0.85} \plotone{S_d_dim.eps}
  \includegraphics[scale=0.32]{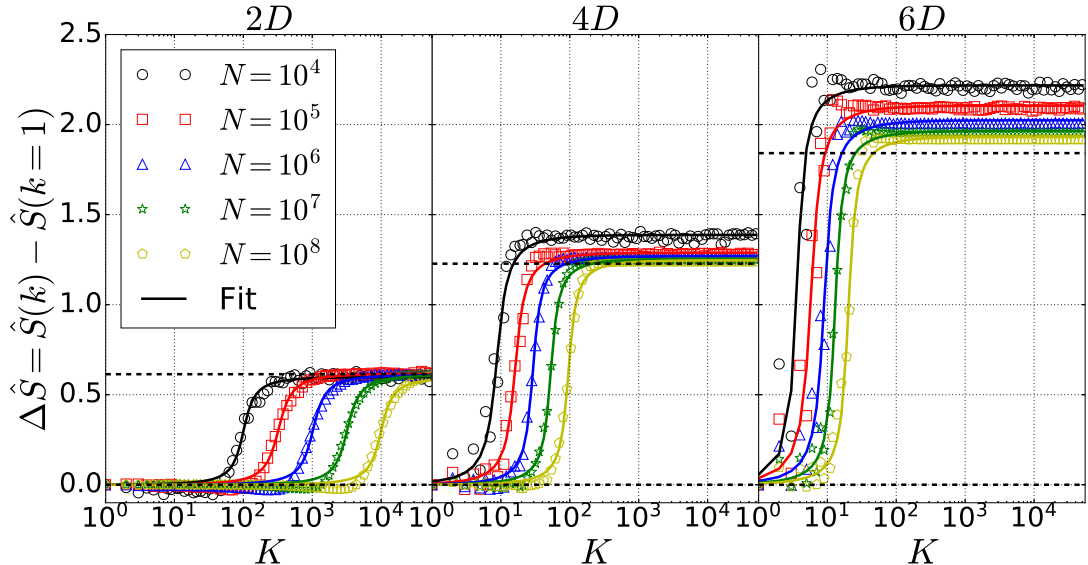}
  \caption{Entropy estimates of samples of $f(x_1,...,x_d)$,
    Eqs.~\eqref{eq:f_d_dim}--\eqref{eq:toy_model}, in $d=2,4,6$
    dimensions shown as a function of the bandwidth $K$. From
    Eqs. \eqref{eq:f_d_dim}--\eqref{eq:S_def}, the theoretical entropy
    value (for $n=1$) is $S_1(k) \approx -d \times 0.307$ for any
    integer $k>0$. The sample reproduces the features of the
    generating function $f$ for $K\lesssim N^{1/d}$ and the entropy
    agrees with the theoretically expected value
    $S(k) = S(k=1) \Rightarrow \Delta \hat{S} = 0$.  For
    $K\gtrsim N^{1/d}$, the sample resembles a uniform distribution
    and the entropy tends to
    $0 \Rightarrow \Delta \hat{S} = d\times 0.307$. These values are
    represented by horizontal dashed lines. Fits (colored solid lines)
    are given by Eq.~\eqref{eq:fit_delta_S}.}
  \label{fig:S_d_dim}
\end{figure*}

The entropy estimates of samples generated with
Eqs.~\eqref{eq:f_d_dim}-\eqref{eq:toy_model} for $n=1$ and ${d=2,4,6}$
are shown in Fig.\ref{fig:S_d_dim}. The black points of the left panel
($2D$) contain the entropy values for the samples shown in
Fig.\ref{fig:coords_S_2D}.
Note that the theoretical entropy value obtained with
Eqs.~\eqref{eq:f_d_dim}-\eqref{eq:S_def} is
$S_{1}(k)\approx -d\times 0.307$, for any integer $k>0$, whereas
$S_{n}(k=0)=0$. Here, $S_{n}$ is the Shannon entropy of the
distribution $f^{(n)}$ defined above. Thus, if the sample is able to
recover the full information of the function $f$ at fixed $k$ value,
we get $\Delta \hat{S}\equiv \hat{S}(k) - \hat{S}(k=1) \simeq 0$, up
to small statistical fluctuations. This is approximately the case for
small $K$, as can be seen in Figs.\ref{fig:coords_S_2D} and
\ref{fig:S_d_dim}. However, for any given number $N$ of points, there
is a critical $K$ value beyond which the function in the continuum $f$
has structures too fine to be resolved or, equivalently, too large a
bandwidth in frequency/Fourier space, for all the information
contained in the function $f$ to be recovered from the discrete
sample, the distribution of which becomes effectively uniform. In this
case, $\hat{S}(k)$ strongly deviates from the entropy of the function
in the continuum, $S(k)$, achieving the maximum, $0$, associated to a
uniform distribution, thus implying
$\Delta \hat{S} = \hat{S}(k) - \hat{S}(k=1)\simeq d\times 0.307$ --
shown as horizontal dashed lines in Fig.~\ref{fig:S_d_dim}.  This
entropy increase is the imprint of the N-S criterion, as
quantitatively demonstrated below.

Let us mention that there is an essential difference between our
entropy estimator $\hat{S}$ and the case covered by the N-S theorem:
whereas the latter refers to a discrete sample of the \textit{actual}
values of the distribution $f$, only \textit{estimates} $\hat{f}_{i}$,
$i=1,\ldots ,N$, of $f$ are available for computing $\hat{S}$. Hence,
the feature of entropy estimators discussed in this work is a (strong)
analogy, a sort of stochastic instance of the original N-S theorem.

The data in Fig.\ref{fig:S_d_dim} can be described by the function
\begin{multline}
  \label{eq:fit_delta_S}
  \Delta \hat{S}(K) = \frac{A}{\pi/2 + \arctan
    \left(BC\right)}\times\\
  \times \left\{\arctan\left[B\left(K - C\right)\right] + \arctan
    \left(BC\right)\right\},
\end{multline}
where $A$, $B$ and $C$ are free parameters representing, respectively,
the maximum entropy increase, the slope of the rising part of the
entropy production curve in Fig.~\ref{fig:S_d_dim} and its delay
(i.e. the $K$ value where the entropy starts to increase). The term
$\arctan \left(BC\right)$ on the right hand side ensures that
${\Delta \hat{S} (K=0)=0}$ and the term
$\pi/2 + \arctan \left(BC\right)$ in the denominator guarantees that
$A = \Delta \hat{S} (K\rightarrow \infty)$. This function -- solid
lines in Fig.\ref{fig:S_d_dim} -- provides reasonable fits for all
values of $d$ and $N$.

We now define $K_{\Delta S/2}$, the value of $K$ at which the entropy
production achieves half of its asymptotic value. Then, with the help
of Eq. \eqref{eq:fit_delta_S} we obtain
\begin{equation}
  \label{eq:k_delta_S}
  K_{\Delta S/2} = \sqrt{B^{-2}+C^2}.
\end{equation}
This quantity represents the critical $K$ value beyond which the
information regarding the continuous function is not adequately
captured by the discrete sample.

\begin{figure}
  \raggedright
  \includegraphics[width=9.0cm]{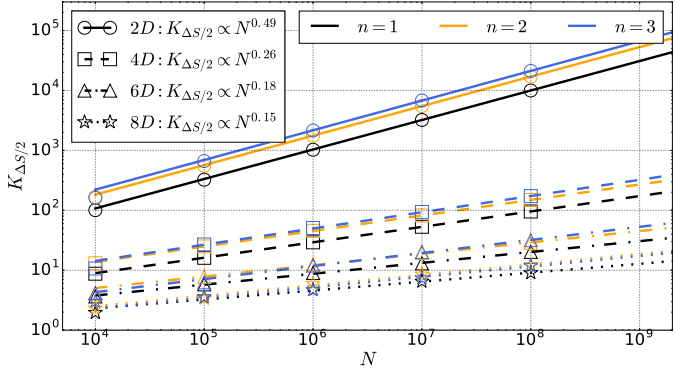}
  \vspace{-0.2cm}
  \caption{Bandwidth $K_{\Delta S/2}$ -- Eq. \eqref{eq:k_delta_S} --
    where the entropy achieves half of its asymptotic value,
    calculated with the fitting values of parameters $B$ and
    $C$. Lines show power-law fits. We obtain approximately
    $K_{\Delta S/2} \propto N^{1/d}$, in agreement with N-S criterion,
    Eq.~\eqref{eq:nyquist}. The same is obtained for different
    functions $f$, i.e. for $n=1,2,3$ in Eq.~\eqref{eq:toy_model}.}
  \label{fig:k_vs_N_d_dim}
\end{figure}

Fig. \ref{fig:k_vs_N_d_dim} shows this quantity, calculated with the
values of $B$ and $C$ obtained in the previous fits for $n=1,2,3$ and
$d=2,4,6,8$. The lines are power law fits to these points. Note that
different values of parameter $n$ represent different models -- see
Eq.~(\ref{eq:toy_model}) -- with ``phase-space'' structures different
from those of Fig.~\ref{fig:coords_S_2D}. For all these different
models we obtain approximately
\begin{equation}
  \label{eq:k_delta_S_2}
  K_{\Delta S/2} \propto N^{1/d},
\end{equation}
in agreement with the N-S criterion, Eq.~\eqref{eq:nyquist}. This
shows that the estimates agree with the entropy of the corresponding
distribution function $f$ in the continuum, if the assumed function
generating the sample fulfills the N-S criterion. This suggests that
the information contents in the sample and the whole function are
approximately equivalent when the criterion is satisfied.

\section{Relaxation of gravitating systems}
\label{sec:relax_dynamical_systems}

Having shown that the entropy estimates agree with the N-S criterion,
i.e. that they capture the information available from a discrete
sample, we now move on to the study of the relaxation process of
finite $N$ gravitating systems. Using the Agama Library
\citep[][]{2018MNRAS.tmp.2556V}, we integrate ensembles of orbits in
the Plummer potential
\begin{equation}
  \label{eq:phi_plummer}
  \phi(r) = -\frac{GM}{a}\frac{1}{\sqrt{1 + (r/a)^2}},
\end{equation}
where $G$ is the gravitational constant, $a$ is a scale radius and $M$
is the total mass. Initial conditions are generated with particles
sampling a top-hat, i.e. a uniform sphere (both in positions and
velocities) of radius $a$ and maximum velocity
${v_{max} = \sqrt{2|\phi(a)|}}$. We set $GM = a = 1$ and integrate the
ensembles for $\approx 300 \tau_{cr}$, with the crossing time
estimated as
${\tau_{cr} = 2\pi \sqrt{\langle r^2\rangle/\langle v^2\rangle}}$,
where these quantities are calculated at $t=0$. The entropy is
estimated with Eqs.~\eqref{eq:S_estimate_0}-\eqref{eq:f_NN}, where
each of the $6$ phase-space coordinates is normalized by its initial
inter-percentile range containing $68\%$ of the data around the
median.

\begin{figure}
  \raggedright
  %\centering
  %\epsscale{0.85} \plotone{S_d_dim.eps}
  %\includegraphics[scale=0.2]{S_plummer.eps}
  \includegraphics[width=9.0cm]{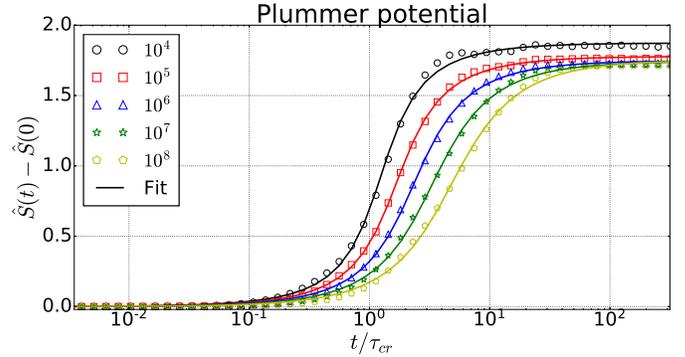}
  \caption{Entropy evolution for an initial uniform sphere (in
    positions and velocities) integrated in a Plummer potential for
    various sample sizes $N$. The resemblance with
    Fig.~\ref{fig:S_d_dim} shows the strength of the analogy of the
    toy model and the evolution of gravitating systems (with $k$
    playing the role of time). Solid lines show fits of
    Eq.~\eqref{eq:fit_delta_S}, replacing $K$ by $t/\tau_{cr}$.}
  \label{fig:S_plummer}
\end{figure}

Fig.\ref{fig:S_plummer} shows the entropy evolution for ensembles of
various sizes $N$ (different colors). We note the resemblance of these
data with that of the toy model, Fig.~\ref{fig:S_d_dim}. Replacing $K$
by $t/\tau_{cr}$, Eq.~\eqref{eq:fit_delta_S} again provides reasonable
fits (solid lines). In accordance with the 2nd law of Thermodynamics,
the time evolution of the system (initially in a non-stationary state)
is such that the entropy increases up to a maximum, where it
stabilizes. In \citetalias{BeS_2018}, we show that the entropy is
conserved for self-consistent (i.e. stationary) samples, also in
agreement with the 2nd law of Thermodynamics.

In the analysis above, we impose a distribution function in the
continuum, i.e. Eqs.~\eqref{eq:f_d_dim}-\eqref{eq:toy_model} for the
toy model and the top-hat initial condition in this section, and ask
ourselves how many data points we need to recover the information
contained in this function. From this point of view, the use of a
finite $N$ limits the possibility of recovering information contained
in fine structures and the entropy increase appears as a result of
information loss (``coarse-graining'') in the entropy estimation.

However, for real gravitational systems such as galaxies and star
clusters, the situation is quite the \textit{opposite}: what is given
a priori, i.e. the real data, is a discrete sample of $N$ stars (or
particles), and the question is if the fine structures developed by
the \emph{assumed} distribution function in the continuum represent
real effects (supported by the discrete data) or rather spurious
features introduced by the theoretical model. Given a discrete sample,
the N-S theorem guarantees a one-to-one correspondence with functions
in the continuum whose characteristic frequencies satisfy the N-S
criterion, Eq.~\eqref{eq:nyquist}. For larger frequencies (finer
structures), many functions in the continuum can be associated to the
same sample and the choice of one specific function (with structures
finer than allowed by the sample) constitutes an \emph{information
  input}, not contained in the sample itself.

Note that the very notion of convergence of a sequence of distribution
functions developing rapidly varying structures (filaments) with the
time evolution is an important conceptual point: such sequences cannot
converge in the point-wise sense and the so-called weak convergence is
a more natural notion in this situation, as pointed out by
\cite{mouhot2011}. This type of convergence means, roughly, that
structures that get arbitrarily fine in the limit must be ``averaged
out'' in order to obtain a well-defined limiting distribution. Our
approach sheds light on this question, by providing a quantitative
criterion to objectively evaluate the ``collapse'' of fine structures
in distributions of particles of macroscopic systems, with fixed
(finite) $N$.

\begin{figure}
  \raggedright
  \includegraphics[width=9.0cm]{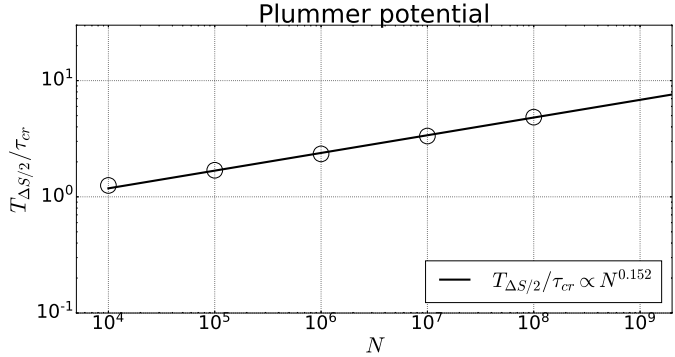}
  \vspace{-0.2cm}
  \caption{Relaxation time, Eq.\eqref{eq:T_delta_S} with the fitting
    values of parameters $B$ and $C$ in Fig.~\ref{fig:S_plummer}, for
    integration in the Plummer potential. The data are well fitted by
    a power law, in connection with the N-S criterion -- see
    Eqs.~\eqref{eq:k_delta_S_2}, \eqref{eq:T_power_law},
    \eqref{eq:K_vs_t}.}
  \label{fig:T_vs_N_plummer}
\end{figure}

Once one recognizes that the real data is a finite $N$ sample (as
opposed to the limit $N\rightarrow\infty$), one can safely consider
the entropy evolution in Fig.~\ref{fig:S_plummer} as characterizing a
real relaxation effect. Analogously to Eq.~\eqref{eq:k_delta_S}, we
define the typical time for this entropy increase as the time when it
achieves half of its asymptotic value:
\begin{equation}
  \label{eq:T_delta_S}
  \frac{T_{\Delta S/2}}{\tau_{cr}} = \sqrt{B^{-2}+C^2}.
\end{equation}
This quantity, calculated with the fitting values of parameters $B$
and $C$, is shown in Fig.\ref{fig:T_vs_N_plummer} (points). This time
scale is well fitted by a power law, which we write as
\begin{equation}
  \label{eq:T_power_law}
  \frac{T_{\Delta S/2}}{\tau_{cr}} \propto N^{\alpha/d},
\end{equation}
where $d=6$ is the dimension of the phase-space. 

The power law obtained for the Plummer potential
--Fig.\ref{fig:T_vs_N_plummer} -- implies $\alpha \approx 0.91$. In
\citetalias{BeS_2018}, the integration of orbit ensembles in an
integrable triaxial potential gives $0.98 \leq\alpha \leq 1.12$,
depending on the initial conditions. Finally, the relaxation times
obtained by \cite{2017JSMTE..04.4001P} in $d=2$ imply $\alpha$ ranging
from $1$ for an integrable system to $\approx 0.3$ for a highly
chaotic one. This weakening of the $N$-dependence of the relaxation
time was interpreted by \cite{2017JSMTE..04.4001P} as a consequence of
an enhancement in the efficiency of phase mixing in the presence of
chaotic motion, and the same effect seems to be present in the results
of \citetalias{BeS_2018}. In light of the results presented in this
work, these outcomes can be interpreted as a direct consequence of the
N-S sampling criterion.

Comparison of Eqs.~\eqref{eq:k_delta_S_2} and \eqref{eq:T_power_law}
indicates that the dynamical evolution of the system is such that the
bandwidth of its distribution function grows with time $t$ as
\begin{equation}
  \label{eq:K_vs_t}
  K \propto \left( \frac{t}{\tau_{cr}}\right)^{1/\alpha}.
\end{equation}
In particular, for $K$ growing linearly in time, as in the simple case
of free particles discussed in \S~\ref{sec:baby_model}, we have
$\alpha = 1$. The results quoted above suggest that an approximately
linear time dependence happens for the evolution in integrable
potentials in general. In such potentials, the use of angle-action
variables $(\vec{\theta},\vec{J})$ allows the reduction of the
dynamics to that of ``free particles'', in which the Hamiltonian
depends only on the momenta, $H = H(\vec{J})$. Although we only have
shown a linear time growth of the bandwidth in the simple case of free
particles considered in \S~\ref{sec:baby_model}, we conjecture that
this could be proven for a large class of integrable potentials, under
suitable technical conditions. Note, in particular, that the harmonic
potential leads to periodic (instead of linear) behavior of the
bandwidth and one can not expect the conjecture to hold true for every
integrable potential.

Concluding, the power-law $N$-dependence for the relaxation time,
Eq.~(\ref{eq:T_power_law}), can be seen as a direct consequence of the
N-S theorem. A linear time growth ($\alpha \approx 1$) of the
bandwidth in frequency/Fourier space of the distribution function for
integrable systems provides a relaxation time scaling as
$\propto N^{1/d}$ for such systems. Moreover, the results of
\citetalias{BeS_2018} show that for a realistic cuspy gravitational
potential hosting large fractions of chaotic orbits, the relaxation
time scale does not seem to change dramatically ($\alpha\approx 0.85$
in that case), at least for the models considered in
\citetalias{BeS_2018}.

\section{Summary}
\label{sec:summary}
To summarize, our results show that the subjective element associated
to the necessity of coarse-graining in order to explain the fast
relaxation of forming or perturbed $N$-body gravitating systems can be
eliminated, via the N-S sampling criterion, if one recognizes as a
fundamental fact that these are finite-$N$ systems whose evolution
saturates their distribution in phase-space at some time determined
essentially by the dimension $d$ and sample size $N$ (and also by the
complexity of trajectories in phase-space). Then, a posteriori one can
look for the distribution function (and for the effective equation
driving its evolution) compatible with the information contained in
the sample at each time. In particular, our results suggest that the
typical relaxation time of integrable systems in a phase-space of
dimension $d$ can be roughly estimated as
$T/\tau_{cr} \propto N^{1/d}$, with the presence of chaotic orbits
accelerating the growth of the bandwidth (i.e. the production of finer
structures) but not dramatically changing the relaxation time
$N$-dependence. Note that this time scale is small, in comparison to
the two-body (collisional) relaxation time, which scales as
$T_{col}/\tau_{cr}\propto N/\ln N$, even for systems containing a
small number of stars, such as open clusters ($N\lesssim 10^4$).

We regard the connection between the N-S criterion (with the
recognition of the discreteness of gravitational systems) and the
entropy evolution shown in this work as a fundamental theoretical
element if one wants to understand the fast collisionless relaxation
of collapsing gravitational structures. Interestingly, these results
seem to be in line with the ``Indispensability of Atomism in Natural
Science'' supported by \cite{book:969643}.

\textit{Acknowledgments:} We thank Jean-Bernard Bru for interesting
discussions and hints. This work has made use of the computing
facilities of the Laboratory of Astroinformatics (IAG/USP,
NAT/Unicsul), whose purchase was made possible by the Brazilian agency
FAPESP (2009/54006-4) and the INCT-A. LBeS is supported by FAPESP
(2014/23751-4 and 2017-01421-0). WdSP is supported by CNPq
(308337/2017-4). MV acknowledges support from HST-AR-13890.001, NSF
award AST-1515001, NASA-ATP award NNX15AK79G. This paper made use of
Agama \citep{2018MNRAS.tmp.2556V}, ANN
\citep{Arya:1998:OAA:293347.293348}, GSL \citep{GSL}, matplotlib
\citep{Hunter:2007}, numpy \citep{Walt:2011:NAS:1957373.1957466} and
scipy \citep{scipy}.

\appendix
Let $d\in \mathbb{N}$ be the phase-space dimension (in a $3D$ space,
$d=6$). We define the probability distribution
$f_{0}:\mathbb{R}^{d/2}\times \mathbb{R}^{d/2}\rightarrow
\mathbb{R}^{+}$ by:
\begin{equation*}
  \label{eq:1}
  f_{0}(\vec{x},\vec{v})\equiv \frac{1}{(2\pi \sigma_{x}\sigma_{v})^{d/2}}\exp \left(-\frac{\vec{x}^{2}}{2\sigma_{x}^{2}}-\frac{\vec{v}^{2}}{2\sigma_{v}^{2}}\right)
\end{equation*}
where $\vec{x},\vec{v}\in \mathbb{R}^{d/2}$ are the particles
positions and velocities, respectively. If no forces act on the
particles (zero-potential) then, at any time $t>0$, the distribution
evolves to:
\begin{equation*}
  \label{eq:5}
  f_{t}(\vec{x},\vec{v})=\frac{1}{(2\pi \sigma_{x}\sigma_{v})^{d/2}}\exp \left[-\frac{(\vec{x}-\vec{v}t)^{2}}{2\sigma_{x}^{2}}-\frac{\vec{v}^{2}}{2\sigma_{v}^{2}}\right].
\end{equation*}

Taking the Fourier transform of $f_{t}$, one gets:
\begin{eqnarray*}
\hat{f}_{t}(\hat{x},\hat{v}) &\equiv &\frac{1}{(2\pi \sigma_{x}\sigma_{v})^{d/2}}\int\mathrm{d}\vec{x}\int\mathrm{d}
\vec{v}\exp \left[-\frac{(\vec{x}-\vec{v}t)^{2}}{2\sigma_{x}^{2}}-\frac{\vec{v}^{2}}{2\sigma_{v}^{2}
}\right]\exp \left[-i(\hat{x}\cdot\vec{x} +\hat{v}\cdot\vec{v}) \right] \\
% &=&\frac{1}{(2\pi \sigma_{x}\sigma_{v})^{d/2}}\int\mathrm{d}
% \vec{x}\int\mathrm{d}\vec{v}\exp \left[-\frac{\vec{x}^{2}}{2\sigma_{x}^{2}
% }-\frac{\vec{v}^{2}}{2\sigma_{v}^{2}}\right]\exp \left[-i\hat{x}\cdot(\vec{x}+\vec{v}t) -i \hat{v}\cdot \vec{v} \right] \\
% &=&\frac{1}{(2\pi
%     \sigma_{x}\sigma_{v})^{d/2}}\int\mathrm{d}\vec{x}\int\mathrm{d}\vec{v}
%     \exp\left[-\frac{\vec{x}^{2}}{2\sigma_{x}^{2}}-\frac{\vec{v}^{2}}{2\sigma_{v}^{2}}\right]
%     \exp\left[-i \hat{x}\cdot\vec{x}-i(\hat{v}+\hat{x}t) \cdot \vec{v}\right] \\
&=&\exp\left[-\frac{\sigma_{x}^{2}\hat{x}^{2}+\sigma_{v}^{2}(\hat{v}+\hat{x}t)^{2}}{2}\right]
    = \exp\left( -\frac{\sigma_v^2}{2}\hat{X}^T {\bf A} \hat{X}\right),
\end{eqnarray*}
where we defined $\hat{X}^T \equiv (\hat{x},\hat{v}) \in \mathbb{R}^d$
and the Hermitian matrix
\begin{equation*}
 % \label{eq:7}
  {\bf A} \equiv\left(
\begin{array}{cc}
t^{2}+s^{2} & t \\ 
t & 1
\end{array}
\right)\text{ },
\end{equation*}
where $s^2\equiv \sigma_x^2/\sigma_v^2$. At this point, note that a
rough estimate of the bandwidth in Fourier space can be obtained as
\begin{eqnarray*}
  K^2(t) &\approx& \langle \hat{x}^2 + \hat{v}^2\rangle = \frac{\int\mathrm{d}\hat{x}^{d/2}\int\mathrm{d}
                   \hat{v}^{d/2}(\hat{x}^{2}+\hat{v}^{2})\hat{f}_{t}(\hat{x},\hat{v})}{\int\mathrm{d}\hat{x}^{d/2}\int\mathrm{d}\hat{v}^{d/2}\hat{f}_{t}(\hat{x},\hat{v})}
         = \frac{\int\mathrm{d}\hat{X}^{d}\hat{X}^2\exp \left( -\frac{\sigma_v^2}{2}\hat{X}^T {\bf A} \hat{X}\right)}{\int\mathrm{d}\hat{X}^{d}\exp \left( -\frac{\sigma_v^2}{2}\hat{X}^T {\bf A} \hat{X}\right)}\\
       % &=& \int\mathrm{d}\hat{x}^{d/2}\int\mathrm{d}\hat{v}^{d/2}(\hat{x}^{2}+\hat{v}^{2})
       %     \exp \left[-\frac{\sigma_{x}^{2}\hat{x}^{2}+\sigma_{v}^{2}(\hat{v}+\hat{x}t)^{2}}{2}\right]
       %     =\int\mathrm{d}\hat{x}^{d/2}\int\mathrm{d}\hat{v}^{d/2}\left[\hat{x}^{2}+(\hat{v}-\hat{x}t)^{2}\right]
       %     \exp\left(-\frac{\sigma_{x}^{2}\hat{x}^{2}+\sigma_{v}^{2}\hat{v}^{2}}{2}\right)\\
       % &=&\frac{1}{(\sigma_{x}\sigma_{v})^{d/2}}\int\mathrm{d}\hat{x}^{d/2}\int\mathrm{d}\hat{v}^{d/2}\left[\sigma_{x}^{-2}(1+t^{2})\hat{x}^{2}+\sigma_{v}^{-2}\hat{v}^{2}\right]\exp\left(-\frac{\hat{x}^{2}+\hat{v}^{2}}{2}\right)
%             = \sqrt{\frac{(2\pi)^d}{\sigma_v^{2d}(\det {\bf A})^{d/2}}}\\
%             &=&\left(\frac{2\pi}{\sigma_{x}\sigma_{v}}\right)^{d/2}\left[\sigma_{x}^{-2}(1+t^{2})+\sigma_{v}^{-2}\right]\text{ }.
        &=& \sigma_{x}^{-2}(1+t^{2})+\sigma_{v}^{-2}.
\end{eqnarray*}
This already shows that, for large times $t$, the bandwidth $K(t)$ is
expected to grow linearly with $t$. A more precise estimate of $K$(t)
involves identifying the ``principal directions'', i.e. a system of
orthogonal axis obtained from $(\hat{x},\hat{v})$ by a rotation such
that one of the new axes points in the direction of largest
dispersion.
% {\bf For this, we write:}
% \begin{eqnarray*}
%   \hat{f}_{t}(\hat{x},\hat{v})
%   &=&\exp \left[-\frac{(\sigma_{v}^{2}t^2+\sigma_{x}^{2})\hat{x}^{2}+2t\sigma_{v}^{2} \hat{x}\cdot\hat{v}+\sigma_{v}^{2}\hat{v}^{2}}{2}\right]\\
%   &=& \prod\limits_{k=1}^{d/2}\exp(-\frac{(\sigma_{v}^{2}t^2+\sigma_{x}^{2})\hat{x}_{k}^{2}+2t\sigma_{v}^{2}\hat{x}_{k}\hat{v}_{k}+\sigma_{v}^{2}\hat{v}_{k}^{2}}{2})
%       =\prod\limits_{k=1}^{d/2}\exp \left[ -\frac{\sigma_{v}^{2}}{2}\left( 
% \begin{array}{cc}
% \hat{x}_{k} & \hat{v}_{k}
% \end{array}
% \right) \left( 
% \begin{array}{cc}
% t^{2}+s^{2} & t \\ 
% t & 1
% \end{array}
% \right) \left( 
% \begin{array}{c}
% \hat{x}_{k} \\ 
% \hat{v}_{k}
% \end{array}
% \right) \right] \text{ },
% \end{eqnarray*}
% where $s^{2}\equiv \sigma_{x}^{2}/\sigma_{v}^{2}$.
For this, we diagonalize the matrix A, concluding that
\begin{equation*}
  \frac{1}{2}\left[ (s^{2}+1)+t^{2}\left( 1\pm
    \sqrt{1+\frac{2(s^{2}+1)}{t^{2}}+\frac{(s^{2}-1)^{2}}{t^{4}}}\right)\right]
\end{equation*}
and
\begin{equation*}
  \label{eq:6}
  \left( 
\begin{array}{c}
  \frac{1}{2t}\left[ (s^{2}-1)+t^{2}\left( 1\pm \sqrt{1+\frac{2(s^{2}+1)}{t^{2}}+\frac{(s^2-1)^2}{t^{4}}}\right) \right]  \\ 
  1
\end{array}
\right)
\end{equation*}
are two eigenvalues and respective orthogonal eigenvectors.
% with the corresponding eigenvalues being
% \begin{equation}
%   \label{eq:8}
% \end{equation}
Expressing the vector
%$\left(\begin{array}{c}\hat{x}_{k} \\ \hat{v}_{k}\end{array}\right)$
%$\left(\hat{x}_{k},\hat{v}_{k}\right)$
$\hat{X}^T = \left(\hat{x},\hat{v}\right)$ in the orthonormal basis associated to
the two eigenvectors above (note that they are not normalized), we
conclude that
% \begin{equation*}
%   \label{eq:9}
% \frac{\sigma_{v}^2}{2}\left( 
% \begin{array}{cc}
% \hat{x}_{k} & \hat{v}_{k}
% \end{array}
% \right) \left( 
% \begin{array}{cc}
% t^2+s^2 & t \\ 
% t & 1
% \end{array}
% \right) \left( 
% \begin{array}{c}
% \hat{x}_{k} \\ 
% \hat{v}_{k}
% \end{array}
% \right) =\frac{\hat{Z}_{k}^{+}(t,\hat{x}_{k},\hat{v}_{k})^{2}}{2\hat{\sigma}_{+,t}^{2}}+\frac{\hat{Z}_{k}^{-}(t,\hat{x}_{k},\hat{v}_{k})^{2}}{2\hat{\sigma}_{-,t}^{2}}\text{ },
% \end{equation*}
\begin{equation*}
\frac{\sigma_v^2}{2}\hat{X}^T {\bf A} \hat{X} = \frac{\hat{Z}^{+}(t,\hat{x},\hat{v})^{2}}{2\hat{\sigma}_{+,t}^{2}}+\frac{\hat{Z}^{-}(t,\hat{x},\hat{v})^{2}}{2\hat{\sigma}_{-,t}^{2}},
\end{equation*}
where
\begin{equation*}
  \label{eq:10}
  \hat{Z}^{\pm}(t,\hat{x},\hat{v})\equiv \frac{\frac{1}{2t}\left[
(s^{2}-1)+t^{2}\left( 1\pm \sqrt{1+\frac{2(s^{2}+1)}{t^{2}}+\frac{
(s^{2}-1)^{2}}{t^{4}}}\right) \right] \hat{x}+\hat{v}}{\sqrt{\frac{1
}{4t^{2}}\left[ (s^{2}-1)+t^{2}\left( 1\pm \sqrt{1+\frac{2(s^{2}+1)}{t^{2}}+
\frac{(s^{2}-1)^{2}}{t^{4}}}\right) +2\right] ^{2}+1}}
\end{equation*}
and
\begin{equation*}
  \label{eq:11}
  \hat{\sigma}_{\pm ,t}^{2}\equiv \frac{1}{\sigma_{v}^{2}}\cdot \frac{2}{
(s^{2}+1)+t^{2}\left( 1\pm \sqrt{1+\frac{2(s^{2}+1)}{t^{2}}+\frac{
(s^{2}-1)^{2}}{t^{4}}}\right) }\text{ }.
\end{equation*}
 Note that $\hat{Z}^{+}(t,\hat{x},\hat{v})$ and
  $\hat{Z}^{-}(t,\hat{x},\hat{v})$ are two vectors in
  $\mathbb{R}^{d/2}$ such that
\begin{equation*}
\hat{Z}^{+}(t,\hat{x},\hat{v})^{2}+\hat{Z}^{-}(t,\hat{x},\hat{v})^{2}=\hat{x}^{2}+\hat{v}^{2}
= \hat{X}^2,
\end{equation*}
and that they point in the ``principal directions'', such that
$\hat{\sigma}_{\pm ,t}^{2}>0$ are the corresponding variances. Observe
also that
\[
\hat{\sigma}_{+,t}\hat{\sigma}_{-,t}=1\text{ }.
\]
This identity is related to the fact that phase-space volume is
preserved by dynamics (Liouville theorem) together with the Parseval
identity for the Fourier transform. At large $t>0$, by a Taylor
expansion, one has:
\begin{equation*}
  t^{2}\left( 1\pm \sqrt{1+\frac{2(s^{2}+1)}{t^{2}}+\frac{(s^{2}-1)^{2}}{
        t^{4}}}\right) =t^{2}\pm \left(
    t^{2}+s^{2}+1-\frac{2s^{2}}{t^{2}}\right) +O(t^{-4}) =
  \left\{
  \begin{array}{l}
    2t^{2}+O(1)\text{ }, \\
    -(s^{2}+1)+\frac{2s^{2}}{t^{2}}+O(t^{-4})\text{ }.
  \end{array}
  \right.
\end{equation*}
% Thus, 
% \begin{eqnarray*}
% t^{2}\left( 1+\sqrt{1+\frac{2(s^{2}+1)}{t^{2}}+\frac{(s^{2}-1)^{2}}{t^{4}}}
% \right)  &=&2t^{2}+O(1)\text{ }, \\
% t^{2}\left( 1-\sqrt{1+\frac{2(s^{2}+1)}{t^{2}}+\frac{(s^{2}-1)^{2}}{t^{4}}}
% \right)  &=&-(s^{2}+1)+\frac{2s^{2}}{t^{2}}+O(t^{-4})\text{ }.
% \end{eqnarray*}
With this we conclude that, at large $t>0$ and fixed $\hat{x},\hat{v}$:

\begin{eqnarray*}
Z^{+}(t,\hat{x},\hat{v}) &=&\hat{x}+O(t^{-1})\text{ }, \\
Z^{-}(t,\hat{x},\hat{v}) &=&\hat{v}+O(t^{-1})\text{ }, \\
\hat{\sigma}_{+,t}^{2} &=&\frac{1}{\sigma_{v}^{2}t^{2}}+O(t^{-4})\text{ },
\\
\hat{\sigma}_{-,t}^{2} &=&\frac{t^{2}}{\sigma_{x}^{2}}+O(1)\text{ }.
\end{eqnarray*}
From this, one obtains the following behavior for the power spectrum
$|\hat{f}_{t}(\hat{x},\hat{v})|^{2}$, at large times:
\[
|\hat{f}_{t}(\hat{x},\hat{v})|^{2}\cong \exp \left[ -\left( \frac{\hat{x}}{
\sigma_{v}^{-1}t^{-1}}\right)^{2}-\left( \frac{\hat{v}}{\sigma_{x}^{-1}t}\right)
^{2}\right] \text{ }.
\]
Thus, the bandwidth $K$, i.e. the largest scales in Fourier space
(smallest scales in real space) of the distribution function,
estimated here as the largest among the dispersions in the directions
$\hat{x}$ and $\hat{v}$, grows as $\sigma_{x}^{-1}t$, for large times
$t>0$. Moreover, in this simple example, the velocity components alone
are responsible for the growth of $K$ (development of fine structures
of $f_{t}(\vec{x},\vec{v})$), for large times.

\bibliography{/Users/lbs/refs_lbs}

\end{document}